\documentstyle[12pt]{article}

\input epsf

\begin{document}
\setlength{\baselineskip}{0.75cm}
\setlength{\parskip}{0.45cm}
\renewcommand{\theequation}{\arabic{equation}}
\renewcommand{\thefootnote}{\fnsymbol{footnote}}
\thispagestyle{empty}
\begin{titlepage}
\begin{flushright}
DO--TH 96/22\\
LMU 03/97\\
hep-ph/9704405
\end{flushright}
\vspace{1cm}
\begin{center}
\large
{\bf Radiatively Corrected Semileptonic Spectra in B Meson Decays}\\
\vspace{1cm}
\normalsize
Changhao Jin$^a$\footnote{E-mail: jin@photon.hep.physik.uni-muenchen.de} 
and Emmanuel A.\ Paschos$^b$\\
\vspace{0.5cm}
{\sl $^a$ Sektion Physik, Universit\"at M\"unchen\\
Theresienstrasse 37, D--80333 M\"unchen, Germany}\\
\vspace{0.3cm}
{\sl $^b$ Institut f\"ur Theoretische Physik, Universit\"at Dortmund\\
D--44221 Dortmund, Germany}
\end{center}
\begin{abstract}
We show how radiative QCD corrections calculated in terms of quarks can be 
incorporated at the hadron level in inclusive semileptonic $B$-meson decays. 
The bound state effects are described by a momentum distribution function of
the $b$ quark.
The summation over the final states and the
averaging over the momentum distribution of the decaying quark render the 
radiative corrections finite.
With this coherent formalism 
we investigate the shape of the electron spectra
for $b\to u$ and $b\to c$ decays as a function of the parameters of the 
theory. 
The resultant
$b\to c$ electron energy spectrum is in agreement with the experimental data.
\end{abstract}
\end{titlepage}
\newpage

\section{Introduction}
Semileptonic $B$-meson decays have been studied for some time now.  
The inclusive decays
\begin{equation}
\bar{B} \rightarrow X_{q}+e+\bar{\nu}_e \, ,
\label{eq:pro1}
\end{equation}
with $\bar{B}$ representing $B^-$ or $\bar{B^0}$ and 
$X_q$ any possible hadronic final state 
containing a charm quark ($q=c$) or an up quark ($q=u$), provide information 
on both couplings $V_{cb}$ and $V_{ub}$, as well as
new information on the internal structure of the $B$-meson.  
In this paper we study the electron energy spectra.
For the theoretical description of the electron spectra 
in eq.(1), strong interactions in
the underlying weak decays must be incorporated, 
since they are responsible for
the confinement of quarks and gluons into hadrons. They will be included in two
steps: as bound state effects and also in the form of gluons radiated during 
the decay.

For inclusive $B$-meson decays,
it was recognized that extended regions of phase space 
involve large values of $q^2$, the momentum transfer squared, 
originating from the large mass of the $B$ 
meson. Consequently the commutator of the two currents 
describing the decay is dominated by distances close to the light--cone.
In this case it is justified to replace the commutator of the weak currents
by their light--cone singularity times a bilocal operator in $b$-quark fields
[1--3].
This replacement together with standard mathematical methods leads to general 
expressions
of the decay spectra which involve a $b$-quark distribution function,
whose origin is non--perturbative.  Thus the semileptonic decays in
eq.(\ref{eq:pro1}) can be described, in direct analogy to deep inelastic 
scattering,
in terms of a quark distribution function, which depends on a new scaling
variable $\xi_+$.

It has also been recognized that the heavy quark 
field can be studied in an effective field theory derived from the 
QCD Lagrangian.  The heavy quark effective theory (HQET) 
sets a framework for keeping track of the heavy quark mass dependence and
for parametrizing nonperturbative phenomena.
The effective theory has been successfully applied to inclusive $B$
decays [4--12]. 
Using the operator product expansion and the method [5--7] of the HQET,
it was possible to derive sum rules for the distribution function, 
which depend on the kinetic energy and the chromomagnetic energy
of the $b$-quark in the $B$-meson.  The numerical values for the sum rules
are determined by static properties of $B$ mesons and 
QCD sum rules \cite{rev}. The sum rules specify the mean value
and the variance of the distribution function.  They imply that the 
distribution function $f(\xi)$ peaks at large
values of $\xi\approx 0.93$ and is very narrow. 

In addition radiative QCD corrections must be included.	
The QCD radiative corrections to the electron energy spectra were 
computed [14--18] 
at the quark level. In the quark decay the phase space ends at the electron
energy $E_e=\frac{m_b}{2}\left( 1-\frac{m_q^2}{m_b^2}\right)$, 
with $m_b$ and $m_q$ the masses of the $b$-quark and the final quark,
respectively.  The quark decay rates as well
as the radiative corrections vanish for 
$E_e>\frac{m_b}{2}(1-\frac{m_q^2}{m_b^2})$, 
whereas in reality the physical endpoint
is $E_e=\frac{M}{2}(1-\frac{M^2_{X_{min}}}{M^2})$ with $M$ the mass of the
$B$-meson and $M_{X_{min}}$ the minimum value of the invariant
mass of the hadronic final state.
In addition, the $O(\alpha_s)$ radiative corrections at the quark level
have logarithmic singularities at
the endpoints.    
For inclusive decays, however, 
we integrate over the phase space of the final 
quark and average over the momentum distribution of the initial quark 
in order to
incorporate the bound state effect.
These steps sum over ensembles of states, render the radiative
corrections finite \cite{KLN,PQS} 
and extend the phase space from the quark level to 
the hadron level.  

In this way we have at our disposal a coherent treatment of perturbative 
and nonperturbative
QCD effects. This treatment has the advantage of accounting correctly for
the phase-space effects and 
producing the electron energy spectra 
which are smooth everywhere up to the physical endpoints.
For instance,
the spectrum is a smooth function of the electron energy 
$E_e$ in the endpoint region
between the $b\to c$ endpoint $E_e=\frac{M}{2}(1-\frac{M_D^2}{M^2})=2.31$ GeV
and the $b\to u$ endpoint $E_e=\frac{M}{2}(1-\frac{M^2_\pi}{M^2})=2.64$ GeV, 
which is useful for extracting
$|V_{ub}|$ since only the $b\rightarrow u$ transition is allowed in this
region.

For the sake of completeness we present in section 2 the general
formalism for inclusive semileptonic B meson decays.  
Then we show in section 3 that the light-cone dominance
leads to an expression for the decay rate in terms
of a distribution function.
In these two sections, we point out how the general formula in eq.(7)
reduces to eq.(24).   
Properties of the distribution function are discussed in section 4 where
they are quantified by the heavy quark effective theory.
In section 5 we propose a method for including  
the QCD radiative corrections to bound states. 
With the above formalism we calculate and analyse, in section 6, 
the electron energy spectra. 

\section{General Formalism}
The inclusive semileptonic decays (\ref{eq:pro1}), in which the $B$ meson
of four-momentum $P$ decays into an electron of four-momentum $k_e$ and
an antineutrino of four-momentum $k_{\nu}$,
are described
by the decay amplitude
\begin{equation}
{\cal{M}} = V_{qb} \frac{G_F}{\sqrt 2} \bar{u}(k_e)\gamma^{\mu}
  (1-\gamma_5)v(k_{\nu}) \langle n|j_{\mu}(0)|B \rangle \,.
\label{eq:amp2}
\end{equation}
Here $V_{qb}$ are the elements of the CKM matrix and $j_{\mu}(x)$
is the weak current, which in terms of quark fields is given by
\begin{equation}
j_{\mu}(x) = \bar{q}(x)\gamma_{\mu}(1-\gamma_5)b(x)
\label{eq:current2}
\end{equation}
and $|B\rangle$ is the B-meson state normalized
according to $\langle B|B\rangle=2P_0(2\pi)^3\delta^3({\bf 0})$.  The basic
quantity for the decay is the second rank tensor
\begin{equation}
W_{\mu\nu} = \sum_n\int \left[ \prod^n_{i=1} \frac{d^3P_i}{(2\pi)^3 
    2E_i}\right]
 (2\pi)^3 \delta^4(P-q-\sum^n_{i=1}P_i) \langle B|j^{\dagger}_{\nu}
  (0)|n \rangle \langle n|j_{\mu}(0)|B\rangle\,,
\label{eq:tensor2}
\end{equation}
where $q$ stands for the four-momentum transferred from the decaying $B$ meson
to the lepton pair, $q=k_e+k_{\nu}$.
It is useful to express the hadronic tensor in terms of a current
commutator
\begin{equation}
W_{\mu\nu} = -\frac{1}{2\pi} \int d^4ye^{iq\cdot y}\langle B|[j_{\mu}(y),
  j^{\dagger}_{\nu}(0)]|B\rangle
\label{eq:comm2}
\end{equation}
because the commutator is more convenient for theoretical considerations.
The hadronic tensor can be decomposed in terms of scalars $W_a(q^2,\, 
q\cdot P)$, $a = 1,\ldots, 5$, as follows :
\begin{equation}
W_{\mu\nu} = -g_{\mu\nu}W_1 + \frac{P_{\mu}P_{\nu}}{M^2} W_2 
 -i\varepsilon_{\mu\nu\alpha\beta} \frac{P^{\alpha}q^{\beta}}{M^2}W_3
         + \frac{q_{\mu}q_{\nu}}{M^2} W_4 
 + \frac{P_{\mu}q_{\nu}+q_{\mu}P_{\nu}}{M^2} W_5 \, .
\label{eq:exp2}
\end{equation}
The tensor $(P_{\mu}q_{\nu}-q_{\mu}P_{\nu})$ 
does not appear because of the time reversal invariance.  We
can express the differential decay rates in terms of the five hadronic
structure functions $W_a,\, a=1,\ldots, 5$.  The decay rate
of the process (\ref{eq:pro1}) in the rest frame of the B meson is
\begin{equation}
\frac{d^3\Gamma}{dE_edq^2dq_0} = \frac{G_F^2|V_{qb}|^2}{16\pi^3 M}
  \left[W_1q^2+W_2(2E_eq_0-2E_e^2-\frac{q^2}{2})+W_3 \frac{q^2}{M}
    (q_0-2E_e) \right] \, .
\label{eq:triple2}
\end{equation}
The structure functions $W_4$ and $W_5$ do not appear
above because their contribution is proportional to the square of
the electron mass and we ignore the lepton masses.
In this general formalism the unknown hadronic structure resides
in the functions $W_a$.  

\section{Light-Cone Dominance}
It is well known  
that integrals like the one in eq.(\ref{eq:comm2}) 
are dominated by distances where
\begin{equation}
0 \leq y^2 \leq \frac{1}{q^2}\,.
\end{equation}
For inclusive semileptonic B-meson decays (\ref{eq:pro1}), 
$q^2$ is timelike and varies
in the physical range 
\begin{equation}
0 \leq q^2 \leq (M - M_{X_{min}})^2\, .
\end{equation}
For extended regions of phase space the momentum transfer squared 
satisfies $q^2 \geq q^2_{ref}$ with $q^2_{ref} \simeq 1 \, {\rm GeV}^2$.
In these regions we expect the dominant contribution to the integral 
in eq.(\ref{eq:comm2}) to come from
distances of the current commutator close to the light-cone.  
The commutator in this region is in fact singular leading to the 
dominant contribution
\begin{equation}
\langle B| \left[ j_{\mu}(y),j_{\nu}^{\dagger}(0) \right] |B\rangle
 = 2(S_{\mu\alpha\nu\beta} -i\varepsilon_{\mu\alpha\nu\beta})
  \left[ \partial^{\alpha}\Delta_q(y) \right] \langle B|\bar{b}(0)
    \gamma^{\beta}(1-\gamma_5)b(y)|B\rangle\, ,
\label{eq:domin3}
\end{equation}
where $S_{\mu\alpha\nu\beta} = g_{\mu\alpha}g_{\nu\beta} + g_{\mu\beta}
 g_{\nu\alpha} - g_{\mu\nu}g_{\alpha\beta}$ 
and $\Delta_q(y)$ is the Pauli--Jordan
function for a free $q$-quark of mass $m_q$.  The factor
in the square bracket in eq.(\ref{eq:domin3}) with the derivative of the 
Pauli-Jordan function has a singularity
on the light-cone.  The last factor with the reduced matrix element
contains the long-distance contribution.  The product of those two
factors is Lorentz
covariant and can be calculated in any Lorentz frame of reference.

The reduced matrix element has a simple Lorentz structure.  It is
in general a function of two scalars $y^2$ and $y\cdot P$
and can be expanded in powers of $y^2$:
\begin{equation}
\langle B|\bar{b}(0)\gamma^{\beta}(1-\gamma_5)b(y)|B\rangle =
  4\pi P^{\beta} \sum_{n=0}^\infty(y^2)^n F_n(y\cdot P)\, .
\label{eq:exp3}
\end{equation}
We shall keep the first term of the series because the higher order 
terms are suppressed by powers of $q^{-2}$.  This approximation is
justified provided the coefficients are not very large.  We can estimate
the coefficients in quark models or the heavy quark effective theory
which indicate that they satisfy
\begin{equation}
(y^2)^n F_n(y\cdot P)\approx e^{-im_{b}v\cdot y}(\Lambda ^2_{QCD}/q^2)^n\, ,
\end{equation}
where $v$ is the velocity of the initial $B$ meson, defined by $v=P/M$.
This behavior motivates the truncation of the series by 
keeping the first term with
$n=0$.

The Fourier transform of $F_0(y\cdot P)$ defines the quark distribution
function
\begin{equation}
f(\xi) = \frac{1}{4\pi M^2}\int d(y\cdot P)e^{i\xi y\cdot P}
 \langle B|\bar{b}(0)P\!\!\!/(1-\gamma_5)b(y)|B\rangle |_{y^2=0}\, .
\label{eq:distr3}
\end{equation}
We can use the inverse Fourier transform
\begin{equation}
F_0(y\cdot P) = \frac{1}{2\pi}\int d\xi 
 e^{-i\xi y\cdot P}f(\xi)
\label{eq:inv3}
\end{equation}
and substitute $F_0$ in eqs.(\ref{eq:exp3}), (\ref{eq:domin3}) and
(\ref{eq:comm2}),  then carry out the
$y$-integration in eq.(\ref{eq:comm2}) and arrive at 
\begin{equation}
W_{\mu\nu} = 4(S_{\mu\alpha\nu\beta}-i\varepsilon_{\mu\alpha\nu\beta})
 \int d\xi f(\xi) \varepsilon (\xi P_0-q_0) \delta
  \left[ (\xi P-q)^2 -m_q^2 \right] (\xi P-q)^{\alpha} P^{\beta} \, .
\label{eq:tensor3}
\end{equation}
The components of the tensor $W_{\mu\nu}$ are expressed in terms of the
distribution function.  We have shown that the dominance of the light--cone
makes possible the expression of the decay rate in terms of a quark
distribution function defined in eq.(\ref{eq:distr3}). 

A special consequence of the decay kinematics is the
occurrence of two roots in the argument of the $\delta$-function in 
eq.(\ref{eq:tensor3}), namely
\begin{equation}
\xi_{\pm}=\frac{q\cdot P \pm \sqrt{(q\cdot P)^2-M^2(q^2-m_q^2)}}{M^2}\, .
\label{eq:root3}
\end{equation}
We shall elaborate on this property below.
The light--cone dominance ascribes the five hadronic structure functions
to a single light--cone distribution function.  The explicit relations
are the following
\begin{eqnarray}
W_1 & = & 2[f(\xi_+) + f(\xi_-)]\, ,\\
W_2 & = & \frac{8}{\xi_+ -\xi_-}[\xi_+f(\xi_+)-\xi_-f(\xi_-)]\, ,\\
W_3 & = & -\frac{4}{\xi_+-\xi_-} [f(\xi_+) -f(\xi_-)]\, ,\\
W_4 & = & 0\, ,\\
W_5 & = & W_3\, .
\end{eqnarray}

The structure functions are 
evaluated in two variables $\xi_{\pm}$.  The second root,
$\xi_-$, is a straightforward consequence of the analysis and corresponds
to the creation of quark--antiquark pairs through the $Z$-diagram because
the energy of the final quark is negative.  The kinematic 
ranges for $\xi_{\pm}$ are 
\begin{eqnarray}
\frac{m_q}{M} & \leq & \xi_+ \leq 1\, ,\\
 -\frac{m_q}{M} & \leq & \xi_- \leq 1 -\frac{m_q}{M}\, .
\end{eqnarray} 
In the light--cone and away from the resonance region $f(\xi_-)$ is
relatively small.  For $b\to c$ decays $\xi_-\stackrel{<}{\sim} 0.75$ where
$f(\xi_-)$ is negligibly small. Scaling of the structure functions with
the scaling variable $\xi_+$ holds when $f(\xi_-)$ is negligible \cite{jp}.  

The expression of the structure functions $W_a$ in terms of a single
distribution function, which depends on two values $\xi_{\pm}$ of the scaling
variable, is
a large simplification.  Substituting the structure functions 
in eq.(\ref{eq:triple2})
we arrive at
\begin{equation}
\frac{d^3\Gamma}{dE_e dq^2 dq_0} =\frac{G_F^2|V_{qb}|^2}{4\pi^3M}\,
  \frac{q_0-E_e}{\sqrt{{\bf q}^2+m_q^2}} \left\{ f(\xi_+)(2\xi_+ E_e M-q^2)
    -(\xi_+ \to \xi_-) \right\}\, .
\label{eq:triple3}
\end{equation}

The remaining unknown function is the reduced matrix element on the light-cone
whose Fourier transform appears as the $b$-quark distribution function.  

\section{Properties of the Distribution Function}
The distribution function obeys positivity and 
is zero for $\xi\leq 0$ or $\xi\geq 1$ \cite{jp}.
Three sum rules for the $b$-quark distribution function are known.
The first one expresses the $b$ quark number conservation \cite{jp}
\begin{equation}
\int_0^1 d\xi f(\xi)=1\, .
\end{equation}
Performing the operator product expansion to reduce the bilocal operator
to local ones and following [5--7] 
in order to expand the matrix element
of the local operator in the HQET, 
two more sum rules were derived.
They determine up to order $(\Lambda_{QCD}/m_b)^2$
the mean value $\mu$ and the variance $\sigma^2$ of the
distribution function, which characterize the position of the maximum and
its width, respectively:
\begin{equation}
\mu \equiv \int_0^1 d\xi\xi f(\xi) = \frac{m_b}{M} (1+E_b)\, ,
\end{equation}
\begin{equation}
\sigma^2 \equiv \int_0^1 d\xi(\xi-\mu)^2 f(\xi) = 
\frac{m_b^2}{M^2} \left( \frac{2K_b}{3} - E_b^2 \right)\, ,
\end{equation}
where 
\begin{equation}
G_b= \frac{1}{2M} \left\langle B\left |\bar{h}_v
\frac{gG_{\alpha\beta}
\sigma^{\alpha\beta}}{4m_b^2} h_v\right |B\right\rangle\,,
\end{equation}
\begin{equation}
K_b= -\frac{1}{2M} \left\langle B\left |\bar{h}_v\,
  \frac{(iD)^2}{2m_b^2}\, h_v\right |B \right\rangle \, ,
\end{equation}    
with $E_b=G_b+K_b$. The first matrix element $G_b$
parametrizes the chromomagnetic
energy arising from the $b$ quark spin and is determined by the mass splitting
between $B^*$ and $B$ mesons [5--7].
For the observed difference $M_{B^*}-M_B = 0.046$ GeV 
\begin{equation}
m_bG_b = -\frac{3}{4}(M_{B^*}-M_B)= -0.034\,\, {\rm GeV}\, .
\end{equation} 
The second matrix element $K_b$ 
parametrizes the kinetic energy of the $b$ quark in the B meson.
It is determined with the help of QCD sum rules and carries a larger error,
leading to the result \cite{ball}
\begin{equation}
2m_b^2K_b=0.5\pm 0.2\,\, {\rm GeV}^2\, .
\end{equation}
Taking $m_b=4.9\pm 0.2$ GeV, the mean value and the variance of the 
distribution function are estimated to be
\begin{equation}
 \mu = 0.93 \pm 0.04 \, ,
\label{eq:constrain1}
\end{equation}
\begin{equation}
\sigma^2 = 0.006 \pm 0.002\, ,
\label{eq:constrain2}
\end{equation}
indicating that the distribution function 
is sharply peaked around its mean value, which is close 
to one.  These 
results are consistent with the original expectations that the distribution
function of a heavy quark peaks at a large value of its argument.

\section{QCD Radiative Corrections}
Special attention must be paid to 
the radiative corrections from the emission of gluons 
and the associated virtual diagrams.  
QCD radiative corrections to the electron energy spectra were studied 
in several
articles [14--18], 
where they were calculated at the quark level. 
As already mentioned in the introduction,
in the application of radiative corrections
at the hadron level we encounter two problems:
the first is to change the quark phase space to the
physical one and the second is the treatment of the logarithmic singularities
to order $\alpha_s$,
which appear at the quark-level endpoints 
$E_e=\frac{m_b}{2}(1-\frac{m_q^2}{m_b^2})$.

These problems may be solved by taking into account the bound state effect.
In the decay of a $B$-meson, the perturbative QCD correction will be modified
by the bound state effects, since QCD confinement implies that free quarks
are not asymptotic states of the theory.  
The bound state effect is described by the $b$-quark distribution function
given in eq.(\ref{eq:distr3}), which is the probability of finding a $b$ quark
with momentum $\xi P$ inside the $B$-meson. 
The substitution of the $b$ quark momentum $p_b$ by $\xi P$ introduces the 
hadronic phase space.
Furthermore, the radiative corrections obtained perturbatively
must be convoluted with the distribution function.
The final contribution for the radiative corrections is given by
\begin{equation}
\frac{d\Gamma_{rad}}{dE_e}=\int^1_{\frac{E_e+\sqrt{E_e^2+m_q^2}}{M}}
d\xi\, f(\xi)\Bigg (\frac{d\Gamma^b_{rad}}{dE_e}\Bigg )_{p_b=\xi P}\, ,
\label{eq:rad5}
\end{equation}
where the quark-level $O(\alpha_s)$ perturbative QCD correction  
$d\Gamma^b_{rad}/dE_e$ was computed analytically in \cite{corbo,jez}.
In this way the endpoints of the perturbative spectra are extented 
from the quark level
to the hadron level and the logarithmic singularities are eliminated. 
As we shall see, the interplay between perturbative and 
nonperturbative QCD effects
is important, especially for the shape of the $b\rightarrow u$ electron
energy spectrum near the endpoint.

We could give at this point formulas for the radiative corrections at the
quark level, but since they are available in two articles \cite{corbo,jez}
we refer to them. The interested reader may consult these articles and use
their formulas for $d\Gamma^b_{rad}/dE_e$ which occurs in eq.(34).
An important property of eq.(34) is that the integral over $\xi$ eliminates
the logarithmic singularities.

In order to 
calculate the decay spectra we need a distribution function $f(\xi)$ 
consistent with the properties of section 4.  We propose the Ansatz
\begin{equation}
f(\xi) = N \frac{\xi (1-\xi)}{(\xi -b)^2+a^2}\theta(\xi) \theta (1-\xi)\, ,
\label{eq:ansatz6}
\end{equation}
where $N$ is the normalization constant and $a$ and $b$ two parameters.  
In case
$a=0$ and $b=m_b/M$, this distribution function reduces to 
a delta function, $\delta(\xi-m_b/M)$, and thus
reproduces the free-quark decay model.  In addition, the constraints
(\ref{eq:constrain1}) and (\ref{eq:constrain2}), stemming from the HQET, 
limit the two constants $a$ and $b$. Other forms of the distribution
function have been proposed in \cite{lk,peter}.

We use eqs.(\ref{eq:triple3}), (\ref{eq:rad5}) and (\ref{eq:ansatz6}) 
to compute the electron energy spectra for $b\to c$
and $b\to u$ decays.  
The calculation is done using (\ref{eq:triple3}) by integrating
first over $q_0$ and then over $q^2$ with the integration limits
\begin{equation}
E_e+\frac{q^2}{4E_e} \leq q_0 \leq \frac{q^2+M^2-M_{X_{min}}^2}{2M}\, ,
\end{equation}
\begin{equation}
0\leq q^2\leq 2E_e \left( M-\frac{M_{X_{min}}^2}{M-2E_e} \right)\, .
\end{equation}
For practical calculations we take $M_{X_{min}}=m_q$ on 
the assumption of quark-hadron duality.

The parameters which enter in the calculation are the final quark mass
$m_q$, the parameters $a$ and $b$ 
(or the equivalent quantities: 
the mean value $\mu$ and the variance $\sigma^2$)
of the distribution function and the
strong coupling constant $\alpha_s$. An important feature is 
the appearance in the decay rates of the physical $B$-meson mass instead of 
the $b$ quark mass.

The necessity of taking into account the interplay 
between perturbative and nonperturbative QCD effects on both $b\to c$ and
$b\to u$ spectra is discussed above. Here we illustrate in Fig.1
that this interplay is important especially in the endpoint region of
the $b\to u$ spectrum. 
The radiative correction calculated with the help of eq.(\ref{eq:rad5}) is 
a smooth function of the electron energy up to the physical endpoint
(solid curve).
The other two curves show
the quark-level perturbative correction without and 
with the Sudakov exponentiation \cite{alt}, respectively.
Radiative corrections without the Sudakov exponentiation 
run off to infinity with increasing energy as expected.
The Sudakov exponentiation eliminates the singularity and gives a decay rate 
finite up to the quark-level endpoint $E_e=\frac{m_b}{2}$.
Beyond this value the correction is zero. For our case, the radiative 
correction, shown in Fig.1,  
remains finite all the way up to the physical endpoint 
$E_e=\frac{M}{2}$.
The property that the averaging over a variable of the initial quark renders
the radiative corrections finite is general and can be applied in other
approaches as well. For example, radiative corrections to the 
analyses \cite{kim,greub} can be easily included using our prescription.
\begin{figure}
\centering
\epsfxsize=15.5cm
\epsffile{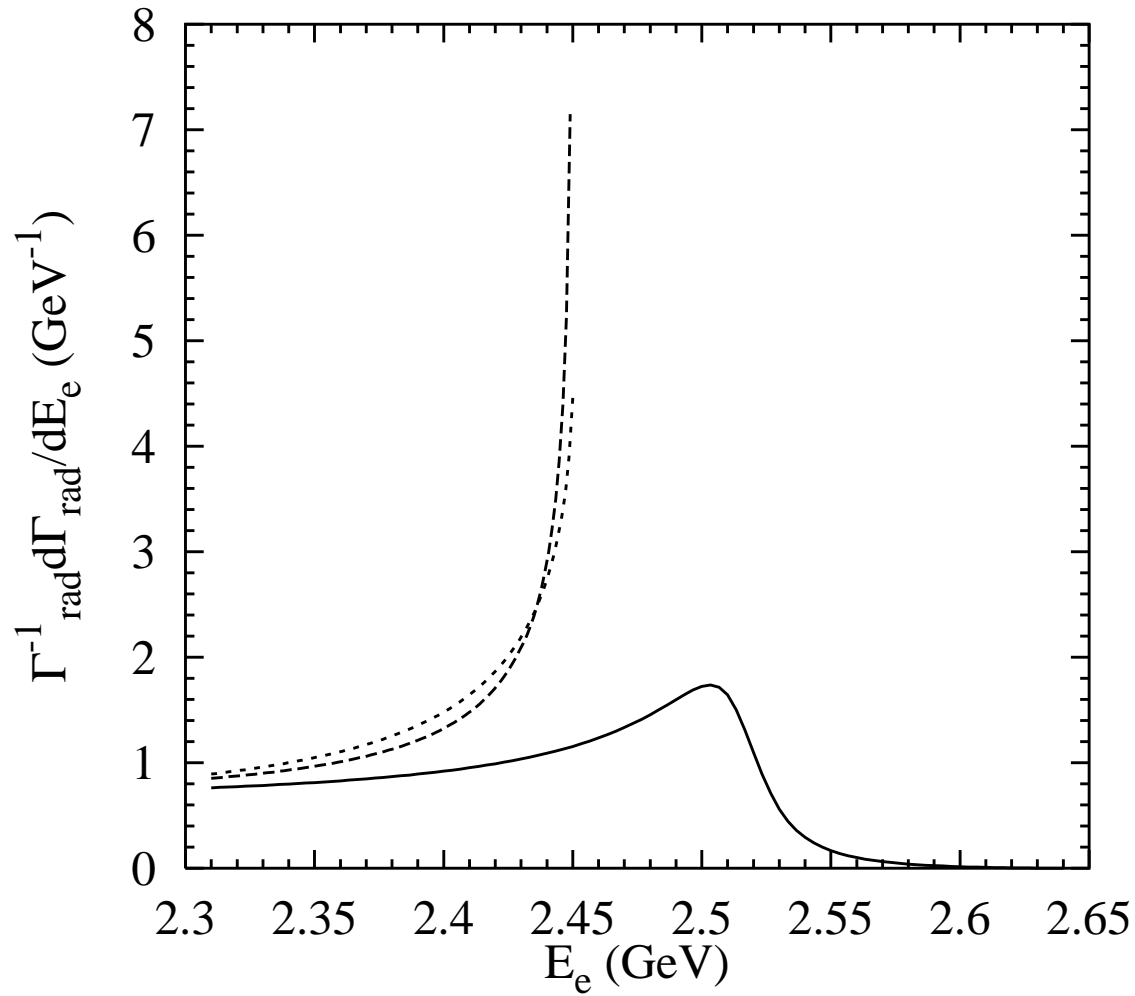}
\vspace{-0.5cm}
\caption{Comparison of the radiative correction to 
the $b\to u$ electron
energy spectrum in the endpoint region as calculated in this paper
(solid line) with the quark-level results of ref.[18]
without (long-dashed line) and with (short-dashed line) the Sudakov
exponentiation for $\alpha_s=0.25, m_b=4.9$ GeV, $m_u=0$,
$a=0.00560$ and $b=0.953$\, .}
\end{figure}

\section{Electron Energy Spectra}
Having at our disposal 
a coherent treatment of perturbative and nonperturbative 
QCD effects,
we study the sensitivity of the shape of the spectrum to various parameters.
As discussed in the previous section, we use eqs.(24), (34) and (35) to 
compute the spectra. More precisely, we integrate eq.(24) over $q_0$ and
$q^2$ and then add the QCD radiative correction from eq.(34) to obtain
the radiatively corrected electron spectrum.
 
The $b\to u$ spectrum is shown in Fig.2 as a function of $\mu$ and $\sigma^2$.
It is evident that the spectrum is much more sensitive
to the mean value $\mu$ of the distribution function than its variance
$\sigma^2$. The shape of the spectrum is insensitive to the values of the mass 
of the final quark $m_u$ and the strong coupling constant $\alpha_s$. 
The effect of the radiative corrections on the spectrum is shown in Fig.3
where we have chosen two values of $\alpha_s=0.25$ (solid curve) and
$\alpha_s=0$ (dashed curve). 
The effect of the radiative correction is moderate.  

\begin{figure}
\centering
\epsfxsize=15.5cm
\epsffile{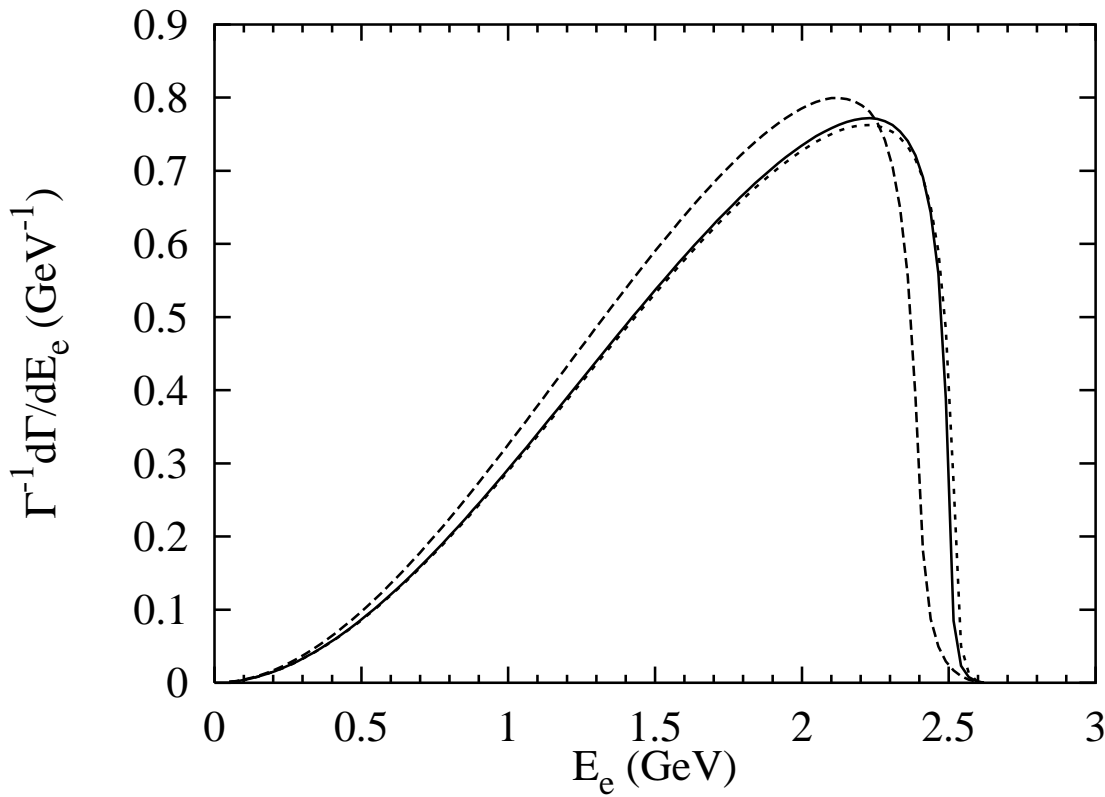}
\vspace{-0.5cm}
\caption{The shape of the electron energy spectrum from 
the $b\rightarrow u$
inclusive semileptonic $B$ meson decay in the rest frame of the $B$ meson
for various values of parameters:
(1) $m_u=0$, $\mu=0.93$, $\sigma^2=0.006$ ($a=0.00560, b=0.953$)
(solid line);
(2) $m_u=0$, $\mu=0.89$, $\sigma^2=0.006$ ($a=0.00992, b=0.913$)
(long-dashed line);
(3) $m_u=0$, $\mu=0.93$, $\sigma^2=0.008$ ($a=0.00679, b=0.960$)
(short-dashed line).
The strong coupling constant is fixed to be 
$\alpha_s=0.25$\, .}
\end{figure}

\begin{figure}
\centering
\epsfxsize=15.5cm
\epsffile{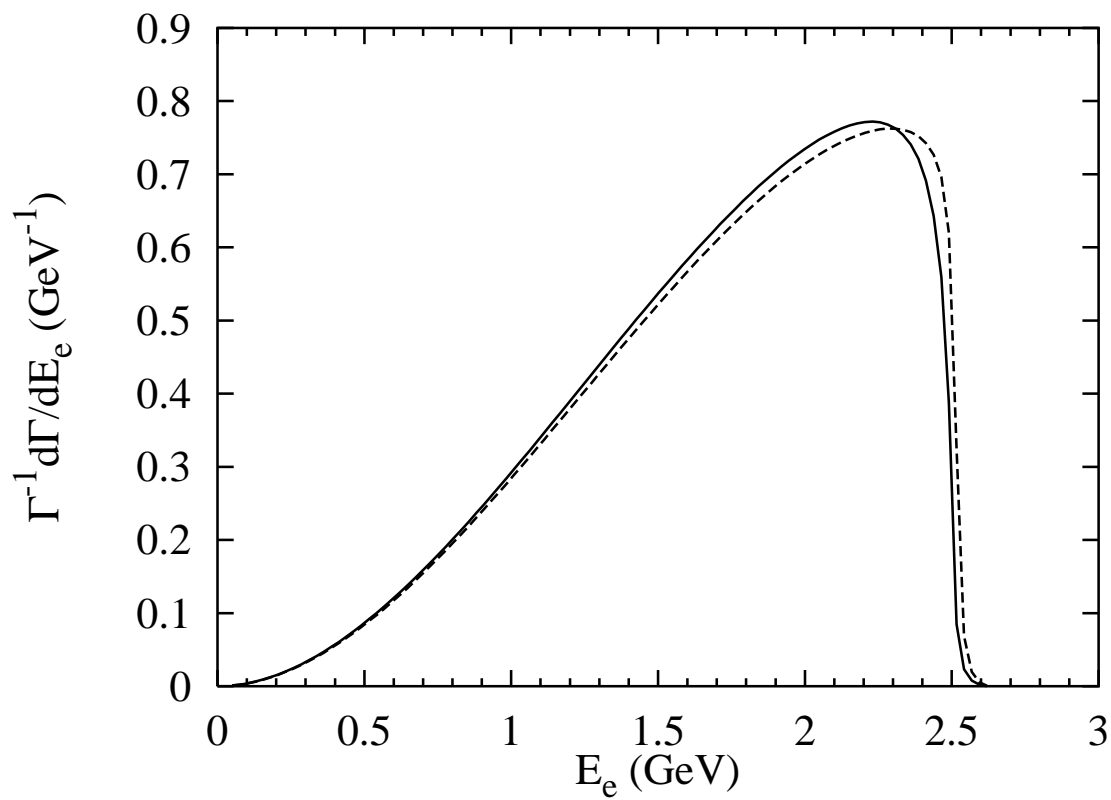}
\vspace{-0.5cm}
\caption{The shape of the electron energy spectrum from 
the $b\rightarrow u$
inclusive semileptonic $B$ meson decay in the rest frame of the $B$ meson
for $m_u=0$, $\mu=0.93$, $\sigma^2=0.006$ ($a=0.00560, b=0.953$),
$\alpha_s=0.25$ (solid line) and $\alpha_s=0$ (dashed line)\, .}
\end{figure}

We study next the dependence of the shape of the electron
energy spectrum for the $b\to c$ decay on various parameters. 
In addition to the previous parameters, the mass of the charm quark plays
now a role.
We show in Fig.4 the electron spectrum as a function of $m_c$, $\mu$
and $\sigma^2$.
The shape of the spectrum is 
a sensitive function of the mass of the charm quark $m_c$ and of 
the mean value $\mu$. It is rather insensitive to the variance $\sigma^2$.  
In Fig.5 we show the electron spectrum with and without radiative corrections.
It is evident that the $b\to c$ spectrum is insensitive to the value of
$\alpha_s$.

\begin{figure}
\centering
\epsfxsize=15.5cm
\epsffile{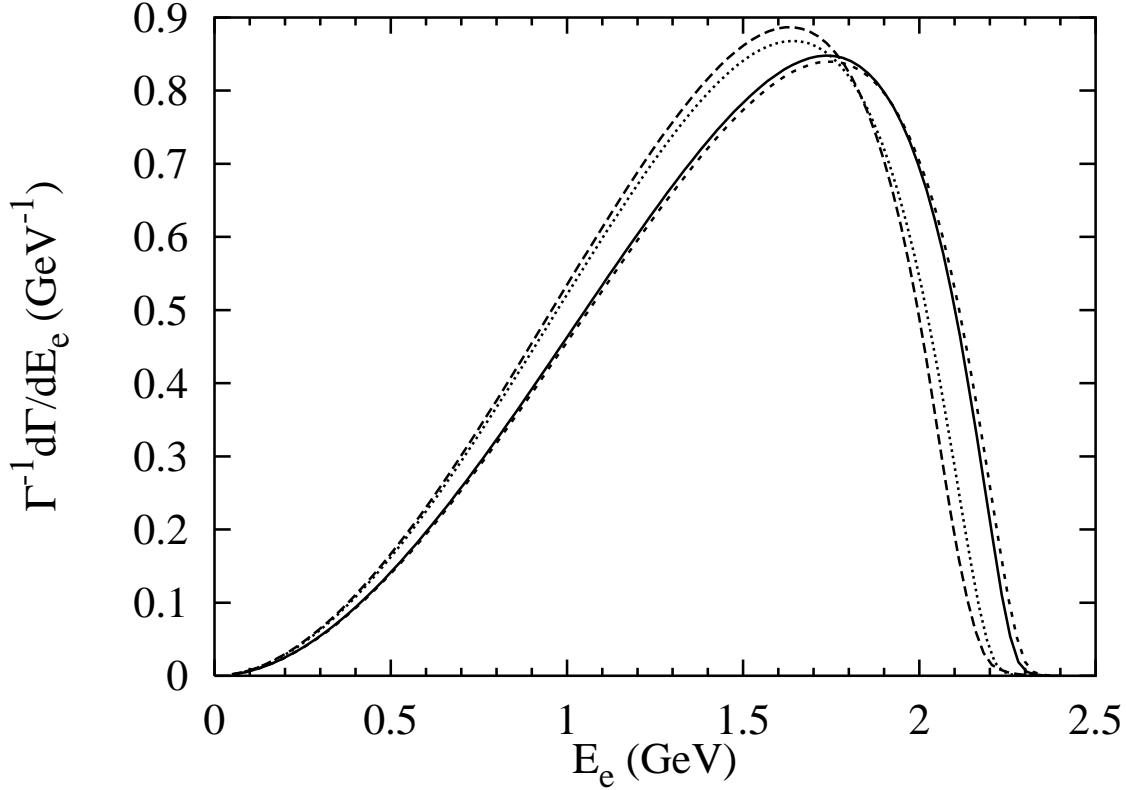}
\vspace{-0.5cm}
\caption{The shape of the electron energy spectrum from 
the $b\rightarrow c$
inclusive semileptonic $B$ meson decay in the rest frame of the $B$ meson
for various values of parameters:
(1) $m_c=1.5$ GeV, $\mu=0.93$, $\sigma^2=0.006$ ($a=0.00560, b=0.953$)
(solid line);
(2) $m_c=1.5$ GeV, $\mu=0.89$, $\sigma^2=0.006$ ($a=0.00992, b=0.913$)
(long-dashed line);
(3) $m_c=1.5$ GeV, $\mu=0.93$, $\sigma^2=0.008$ ($a=0.00679, b=0.960$)
(short-dashed line);
(4) $m_c=1.7$ GeV, $\mu=0.93$, $\sigma^2=0.006$ ($a=0.00560, b=0.953$)
(dotted line).
The strong coupling constant is fixed to be $\alpha_s=0.25$\, .}
\end{figure}

\begin{figure}
\centering
\epsfxsize=15.5cm
\epsffile{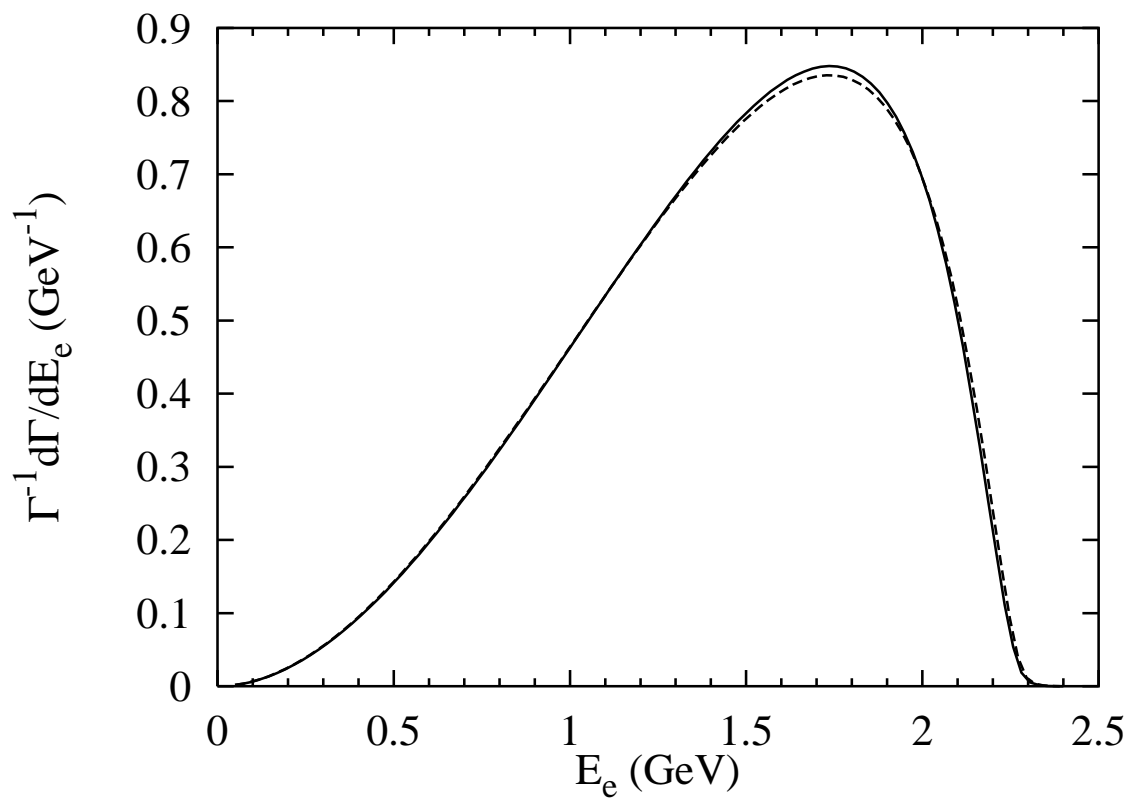}
\vspace{-0.5cm}
\caption{The shape of the electron energy spectrum from 
the $b\rightarrow c$
inclusive semileptonic $B$ meson decay in the rest frame of the $B$ meson
for $m_c=1.5$ GeV, $\mu=0.93$, 
$\sigma^2=0.006$ ($a=0.00560, b=0.953$),
$\alpha_s=0.25$ (solid line) and $\alpha_s=0$ (dashed line)\, .}
\end{figure}

With the parameters determined so far we can calculate 
the $b\to c$ spectrum and compare it
with the recent experimental data from the CLEO 
collaboration \cite{CLEO}. We present the result in Fig.6, 
where the theoretical curve has been boosted to the
rest frame of the $\Upsilon(4S)$ resonance. 
The values of $a=0.0118$ and $b=0.931$ are still consistent with the sum
rules in eqs.(\ref{eq:constrain1}) and (\ref{eq:constrain2}).
This is a direct calculation of the spectrum and not a $\chi^2$ fit.
The agreement with the experimental data is good. 

\begin{figure}
\centering
\epsfxsize=15.5cm
\epsffile{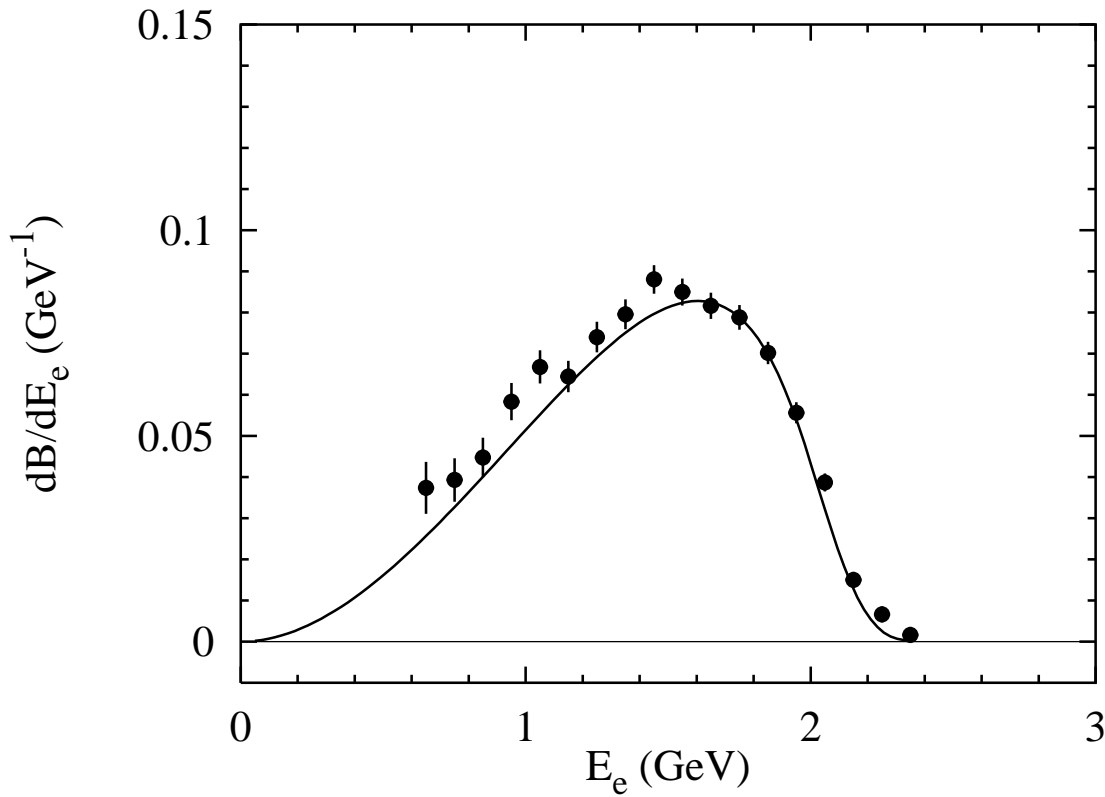}
\vspace{-0.5cm}
\caption{The predicted electron energy spectrum 
compared with the CLEO data.
The theoretical calculation uses $m_c=1.61$ GeV, $a=0.0118$, $b=0.931$,
$\alpha_s=0.25$ and $\beta =0.061$ for the velocity of the $B$ meson.
The spectrum is normalized to the $B$-meson semileptonic branching fraction.}
\end{figure}

The above analyses indicate the sensitivity of the spectral shape to various
parameters and  
imply that
a detailed fit\footnote{Such a fit should also 
account for detector resolution and bremsstrahlung.}
to the measured spectrum can impose strong constraints on 
the mean value of the distribution function and the mass of the charm quark.
This procedure will reduce the theoretical uncertainty 
in the calculation of the
semileptonic decay width of the $B$-meson \cite{width} and 
improve the accuracy of the predictions for the $b\to u$ spectrum.
Precise determinations of $|V_{cb}|$ and $|V_{ub}|$ may be gained 
from inclusive semileptonic $B$-meson decays.
Finally, the same tensor structure appears 
in the decay $B\to J/\psi+X$ \cite{soldan} and a universal fit of both
processes would be of interest.
Dedicated studies of the inclusive B decays will also offer more insight into
the internal structure of hadrons. 
\bigskip
\bigskip
\begin{flushleft}
{\bf Acknowledgements}
\end{flushleft}
We would like to thank R. Poling and R. Wang for kindly providing us
with their data on inclusive B decays and P. Soldan for reading the paper.
The financial support of the Bundesministerium f\"ur Bildung und Forschung
under contract No.~056DO93P(6) is gratefully acknowledged.

\end{document}